\documentclass[aps,prb,showpacs,twocolumn,amsmath,amssymb,superscriptaddress]{revtex4}
\usepackage{graphicx}
\usepackage{dcolumn}
\usepackage{color}
\usepackage{bm}
\usepackage[normalem]{ulem}

\begin{document}

\title{Pressure effects on crystal and electronic structure of bismuth tellurohalides}

\author{I.~P.~Rusinov}
\affiliation{Tomsk State University, pr. Lenina, 36, Tomsk, 634050 Russia}
\affiliation{St. Petersburg State University, Universitetskaya nab., 7/9, St. Petersburg, 199034 Russia}

\author{T.~V.~Menshchikova}
\affiliation{Tomsk State University, pr. Lenina, 36, Tomsk, 634050 Russia}

\author{I.~Yu.~Sklyadneva}
\affiliation{Tomsk State University, pr. Lenina, 36, Tomsk, 634050
Russia} \affiliation{Karlsruher Institut f\"ur Technologie, Institut
f\"ur Festk\"orperphysik, D-76021 Karlsruhe, Germany}
\affiliation{Donostia International Physics Center (DIPC), Paseo de
Manuel Lardizabal, 4, 20018 San Sebasti\'an/Donostia, Basque
Country, Spain} \affiliation{Institute of Strength Physics and
Materials Science, pr. Academicheskii 2/1,634021, Tomsk, Russian
Federation}

\author{R.~Heid}
\affiliation{Karlsruher Institut f\"ur Technologie, Institut f\"ur Festk\"orperphysik, D-76021 Karlsruhe, Germany}

\author{K.-P.~Bohnen}
\affiliation{Karlsruher Institut f\"ur Technologie, Institut f\"ur Festk\"orperphysik, D-76021 Karlsruhe, Germany}

\author{E.~V.~Chulkov}
\affiliation{Tomsk State University, pr. Lenina, 36, Tomsk, 634050 Russia}
\affiliation{St. Petersburg State University, Universitetskaya nab., 7/9, St. Petersburg, 199034 Russia}
\affiliation{Donostia International Physics Center (DIPC), Paseo de Manuel Lardizabal, 4, 20018 San Sebasti\'an/Donostia, Basque Country, Spain}
\affiliation{Departamento de F\'isica de Materiales, Facultad de Ciencias Qu\'imicas, UPV/EHU, 20080 San Sebasti\'an/Donostia, Basque Country, Spain}
\affiliation{Centro de F\'{i}sica de Materiales CFM--MPC, Centro Mixto CSIC--UPV/EHU, 20080 San Sebasti\'an/Donostia, Basque Country, Spain}

\begin{abstract}
We study the possibility of pressure-induced
transitions from a normal semiconductor to a topological insulator (TI)
in bismuth tellurohalides using density functional theory and
tight-binding method. In BiTeI this transition is
realized through the formation of an intermediate phase, a Weyl
semimetal, that leads to modification of surface state dispersions.
In the topologically trivial phase, the surface states exhibit a
Bychkov-Rashba type dispersion. The Weyl semimetal phase exists in a narrow
pressure interval of 0.2 GPa. After the Weyl semimetal--TI transition occurs,
the surface electronic structure is characterized by gapless states
with linear dispersion.  The peculiarities of the surface states modification
under pressure depend on the band-bending effect.
We have also calculated the frequencies of Raman
active modes for BiTeI in the proposed high-pressure crystal phases
in order to compare them with available experimental data.
Unlike BiTeI, in BiTeBr and BiTeCl the topological phase transition
does not occur. In BiTeBr, the crystal structure changes with
pressure but the phase remains a trivial one. However, the
transition appears to be possible if the low-pressure crystal
structure is retained. In BiTeCl under pressure, the
topological phase does not appear up to 18 GPa due to a relatively
large band gap width in this compound.
\end{abstract}

\date{\today}

\pacs{73.43.Nq, 73.20.At, 78.40.Kc, 78.30.Am, 62.50.-p}

\maketitle

\section{INTRODUCTION}

Materials with strong spin-orbit coupling (SOC) open up exciting
possibilities in the rapidly developing area of solid state physics
-- spintronics. Such perspective materials are, for example,
topological insulators (TI) which simultaneously combine the
properties of a semiconductor in the bulk and a metal on the
surface.\cite{pankratov_1987,fu_2007,xia_2009,eremeev_2010,eremeev_landolt_2012}
The metallic behavior of the surface is caused by the presence of
special spin-polarized surface states with Dirac-type dispersions
which are topologically protected from backscattering. Other
promising candidates for spintronics are bismuth tellurohalides
(BiTe$X$, $X$=I,Br,Cl). These non-centrosymmetric compounds are topologically trivial
but characterized by a giant Bychkov-Rashba-type spin
splitting~\cite{BR} of bulk and surface electronic
bands.\cite{ishizaka_2011,bahramy_2011,eremeev_2012,eremeev_jetpl_2012,landolt_2012,crepaldi_2012,rusinov_2013,sakano_2013,eremeev_bitebr}.
Both topological insulators and bismuth tellurohalides can be used
for designing new spintronic and  magnetoelectric devices such as
spin transistors~\cite{dolcini,krueckl,datta-das,datta-das2} as well
as for the creation of quantum computers.

Recently it was predicted that BiTeI transforms from a trivial phase
into a topological insulator by applying an external pressure of
1.7--4.1 GPa.\cite{bahramy2_2012,bahramy_2012} The topological phase
transition (TPT) was also observed experimentally at 2--2.9~GPa and
3.5~GPa using infrared spectroscopy~\cite{xi2013} and Shubnikov-de
Haas oscillations measurements,\cite{ideue2014} respectively.
In Ref.~\citenum{tran2014}, within optical measurements the TPT doesn't observed.
In the theoretical study within density functional theory (DFT) the TPT was
found at 4.5~GPa.\cite{chen_JPCC} Using \textit{ab initio} based
tight-binding (TB) calculations it was shown~\cite{vanderbilt2014} that the
TPT is accompanied by the formation of an intermediate phase,
a Weyl semimetal, which is characterized by one or more pairs of
band-touching points (Weyl nodes) between valence and
conduction bands. The possibility
of topological phase transition in the related tellurohalides,
BiTeBr and BiTeCl, has not been studied yet.

To draw a conclusion about the existence of pressure-induced
topological phases in these compounds, one should first find out
whether any pressure-induced crystal phase transitions (CPT) would
occur under pressure. In the case of BiTeI, the experimental
observation of x-ray diffraction~\cite{xi2013,chen_JPCC} and
Raman~\cite{ponosov2013,tran2014} spectra reveals the CPT at
pressure of $\sim$8--9~GPa which is by factor of 2-3 higher then
the pressure of TPT. In Ref.~\onlinecite{chen_JPCC}, the
orthorhombic Pnma structure was proposed as a high-pressure phase by
comparing the DFT-obtained enthalpy for the low-pressure phase and
Pnma. It was also shown that this hexagonal--orthorhombic CPT occurs
at a pressure of $\sim$6~GPa.\cite{chen_JPCC} For the BiTeBr
compound, the CPT has been experimentally observed at similar
pressures (6--7~GPa).\cite{manjon_2016}

A thorough investigation of the topological transition in BiTeI
requires a careful consideration of surface electronic properties.
The fact is that in the bismuth tellurohalides, the band-bending is
of special significance because this effect induces additional
spin-polarized surface states which have been observed both in angle
resolved photoemission spectroscopy (ARPES) measurements and in DFT
calculations~\cite{eremeev_2012,landolt_2012,crepaldi_2012,sakano_2013}.
The band-bending arises from the polar nature of the compounds and
is caused by a charge redistribution at the surface--vacuum
boundary. The redistribution changes the effective potential level
in the surface region relatively to the bulk and, thereby, shifts
the chemical potential near the surface. Till now the surface
electronic structure of BiTeI under pressure has been investigated
without taking into account the band-bending.

\begin{figure*}[ht!]
\centering
\includegraphics[width=2\columnwidth]{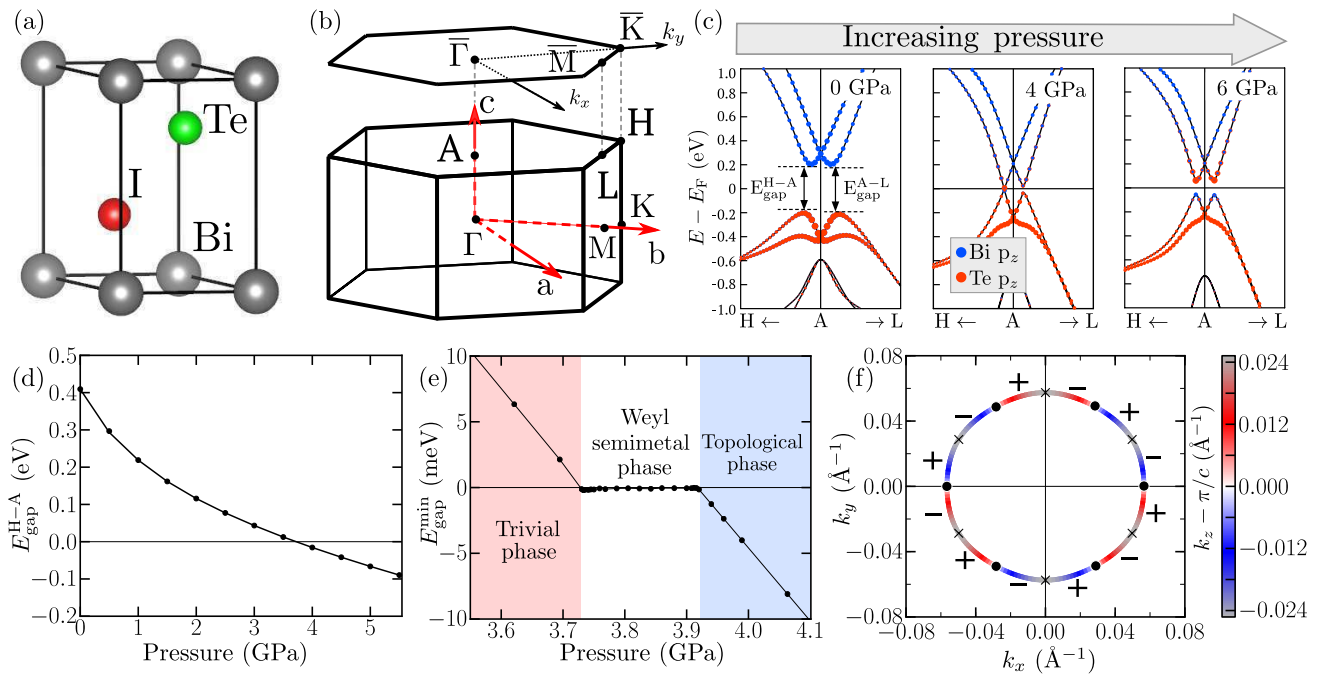}
\caption{(a) A unit cell of BiTeI in the P3m1 hexagonal crystal
structure. (b) Bulk and surface Brillouin zones (BZ). (c) Evolution
of bulk band dispersions near the Fermi level along the H-A-L line
of the BZ with pressure. The size of blue (red) circles reflects the
contribution of bismuth (tellurium) $p_{z}$ states to the electronic
bands. (d) $E^{\mathrm{H-A}}_{\mathrm{gap}}$ and (e)
$E^{\mathrm{min}}_{\mathrm{gap}}$ as a function of pressure near the
topological phase transition. (f) Trajectories of Weyl points
(projected onto the ($k_{x},k_{y}$) plane) during the transition
from a trivial phase to the topological insulator. Colors reflect a
shift of the Weyl nodes along the $k_z$ direction (the right panel).
Dots and crosses show the position of Weyl points at a pressure
$P_{c}\sim$3.73~GPa and at the annihilation point, respectively.
The chirality of Weyl points which move clockwise and
anticlockwise is shown by the signs plus and minus.}
\label{fig:structure}
\end{figure*}

\begin{figure}[bh!]
\includegraphics[width=0.95\columnwidth]{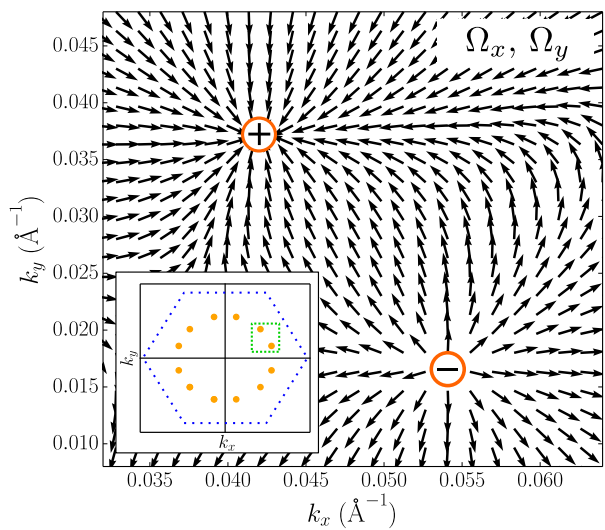}
\caption{In-plane components of the Berry curvature
vector field ($\Omega_{x}$, $\Omega_{y}$)
for BiTeI under pressure of 3.84~GPa in the ($k_{x},k_{y}$) plane at
$k_z=\pi/c-0.012$~\AA$^{-1}$. The insert shows schematically the
positions of all Weyl points. The BZ region demonstrated in the main
figure is marked by a green square.} \label{fig:berry}
\end{figure}

\begin{figure*}[ht!]
\centering
\includegraphics[width=2\columnwidth]{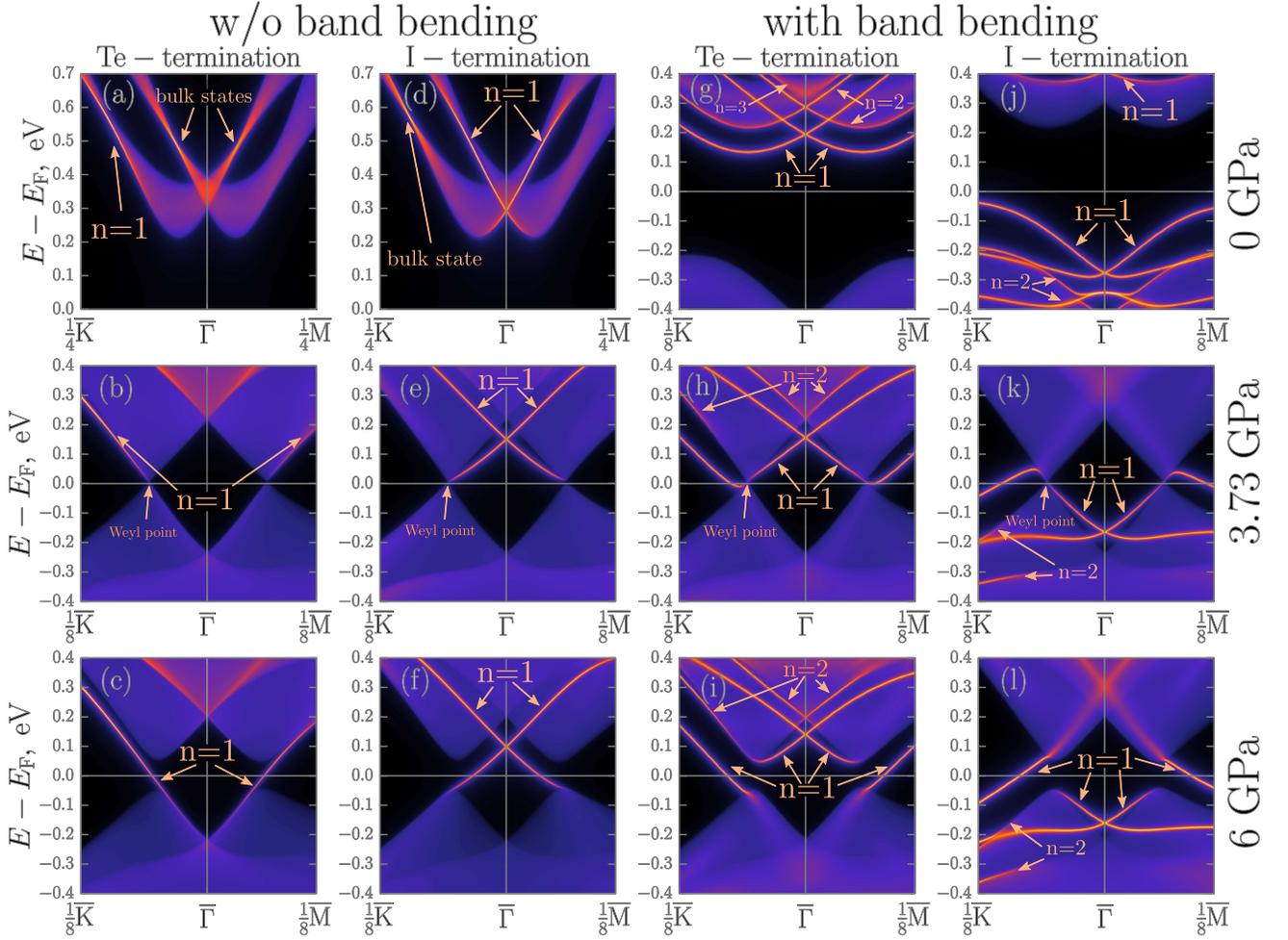}
\caption{Surface band structure of BiTeI calculated for the
Te-terminated surface at (a) 0~GPa, (b) 3.73~GPa, and (c) 6~GPa
without considering the band-bending. (d--f) The same as in (a--c)
for the I-terminated surface. (g-l) The same as in (a-f), but with
taking into account the band-bending. The electronic states
localized in the topmost, second and third trilayers are labeled by
$n=1$, $n=2$ and $n=3$, respectively. The positions of bulk Weyl
points at a pressure of 3.73 GPa are shown by arrows.}
\label{fig:surf}
\end{figure*}

\begin{figure}[ht!]
\centering
\includegraphics[width=\columnwidth]{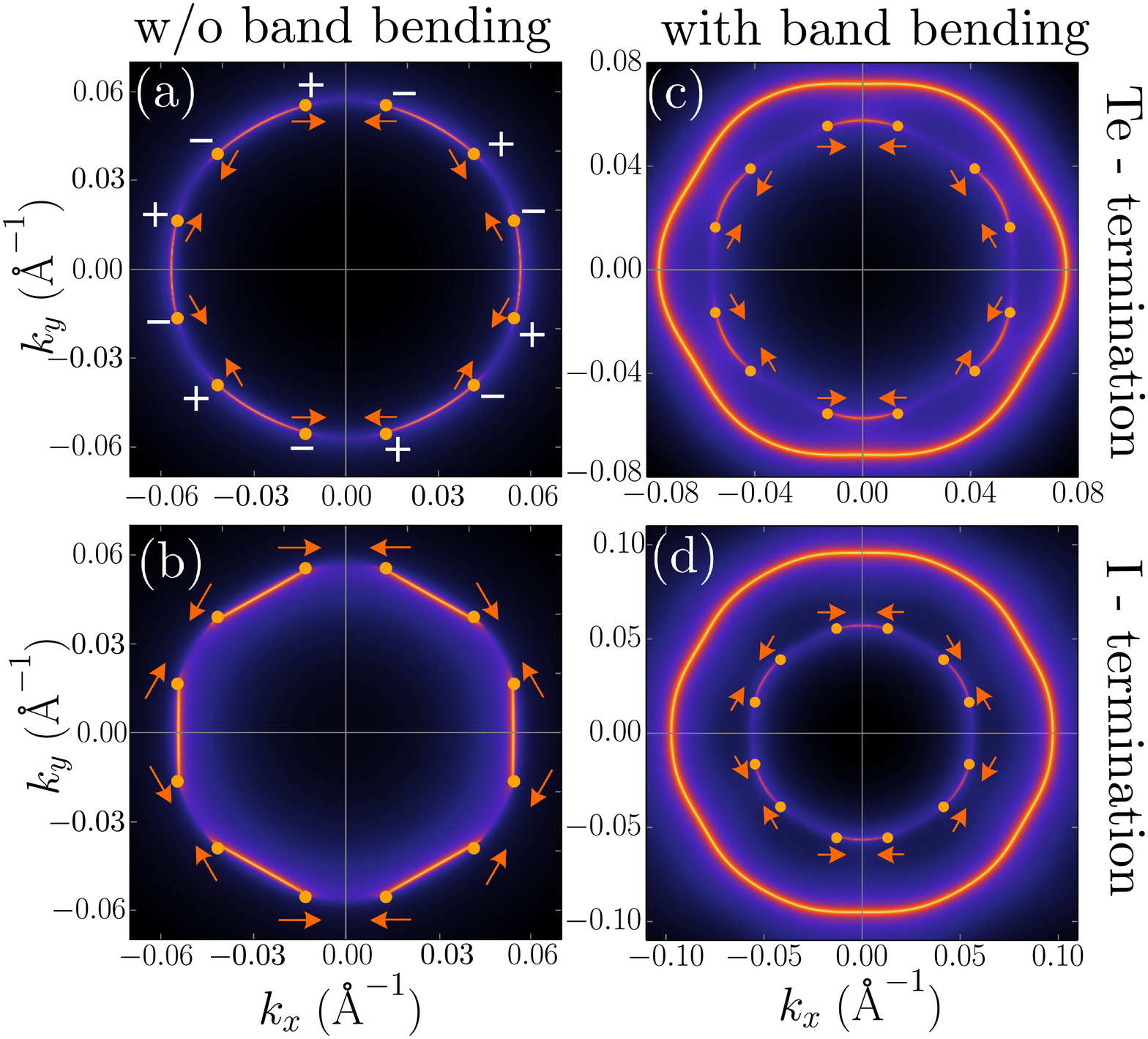}
\caption{Constant energy contours calculated
at the Fermi level of the BiTeI (001) surface at a pressure of
3.84~GPa. The positions of Weyl points are denoted by orange dots.
Clockwise and counterclockwise displacements of the Weyl points with
increasing pressure are shown by arrows. (a,b) The case of
Te-terminated (a) and I-terminated (b) surfaces without taking
account of the band-bending. (c,d) The same with account of the
band-bending. On panel (a) the chirality of Weyl points is
shown.} \label{fig:fermi_arcs}
\end{figure}

Here we present a theoretical study both of the topological and crystal
phase transitions in the bismuth tellurohalides. In the case of
BiTeI, we demonstrate the features of bulk and surface electronic
spectra in the topologically trivial phase, in the case of Weyl
semimetal and in the topological insulator. It is found that the
intermediate phase in  BiTeI, a Weyl semimetal, appears in the
pressure range of 3.7--3.9~GPa which can be experimentally detected.
It is also shown that the effect of band-bending plays a crucial
role in the surface electronic structure formation. By tracing the
modification of surface electronic spectra with pressure the
mechanism of changing the dispersion of surface states is revealed.
An analysis of BiTeBr and BiTeCl shows the absence of TPT in these
compounds. In BiTeBr, a crystal phase
transition occurs before the TPT that precludes the latter. In
BiTeCl, the topological phase does not appear due to a relatively
large band gap at zero pressure.

\section{RESULTS AND DISCUSSION}

\subsection{COMPUTATIONAL DETAILS}

The bulk electronic structure calculations were carried out within DFT
using the projector augmented-wave (PAW) method realized in the VASP
code and the PBE exchange-correlation functional~\cite{pbe}.
The spin-orbit coupling was accounted by a second-variation
method. Crystal lattice parameters and atomic positions were optimized for
pressures up to 8, 10, 18~GPa for BiTeI, BiTeBr and BiTeCl, respectively,
using PBE exchange-correlation functional. Under the structural relaxation
the crystal symmetry kept the same. The optimization was performed for
all structures of BiTeX (X=I, Cl, Br) considered in the paper. At zero pressure,
for BiTeI and BiTeCl compounds overstimation of $a$ and $c$ parameters is 2\% and 7\%, 
respectively. For BiTeBr compound~---2\% and 6\%, respectively.

For simulation surface under pressure, the large slab has been
constructed on the basis relaxed parameters of bulk structure under
the pressure. The large slab Hamiltonian derived from the
bulk one is then used to calculate surface Green functions.~\cite{Sancho:84,Sancho:85,Henk:93}
So, tight-binding models were constructed using
WANNIER90 code.\cite{marzari,WZhang}. The chosen basis consists of six
spinor $p$-type orbitals for each atom: $\left| p^{\uparrow}_{x}
\right>$, $\left| p^{\uparrow}_{y} \right>$, $\left|
p^{\uparrow}_{z} \right>$, $\left| p^{\downarrow}_{x} \right>$,
$\left| p^{\downarrow}_{y} \right>$, $\left| p^{\downarrow}_{z}
\right>$. The low-lying $s$ orbitals are not taken into
consideration. To study the bulk electronic spectra near the point
of TPT two tight-binding Hamiltonians are constructed: one for the
topologically trivial phase of BiTeI and the other for the TI phase,
$\hat{H}_{\mathrm{triv}}$ and $\hat{H}_{\mathrm{top}}$,
respectively. For each intermediate pressure the Hamiltonian is
taken as a linear combination: $\hat{H}=\eta
\hat{H}_{\mathrm{top}}+(1-\eta)\hat{H}_{\mathrm{triv}}$. Here
$\eta=0$ and $\eta=1$~ correspond to a pressure of 3.6~GPa and
4~GPa, respectively.

The band-bending effect was accounted by shifting the on-site matrix elements of surface atoms. The value of shifting is determined by the potential gradient
obtained in the first-principles calculations for slabs of BiTeX (X=I, Cl, Br) at zero pressure.\cite{eremeev_2012,eremeev_bitebr} These values of band bending shift were applied for all surfaces of BiTeX (X=I, Cl, Br)  under pressure.

Dynamical properties of the bulk bismuth tellurohalides were
calculated within the density-functional perturbation
theory\cite{zein} in a mixed-basis pseudopotential
approach.\cite{Louie:79,Meyer,Heid:99}. The details of the
calculation as well as the spin-orbit coupling implementation within
the mixed-basis pseudopotential method can be found in
Ref.~\onlinecite{Heid:10,Sklyadneva}.

\subsection{Bulk band structure of BiTeI}

BiTeI crystalizes in a hexagonal structure, P3m1
(Fig.~\ref{fig:structure}(a)).
At zero pressure, the compound is a
semiconductor.
Using fully optimized crystal structure parameters
($a$, $c$ and atomic coordinates), we obtained the band
gap width of 408~meV which is in good agreement both
with experimental measurements (380 meV~\cite{ishizaka_2011}) and 
$GW$-calculations (400~meV in the  LDA+$GW$ scheme~\cite{rusinov_2013}).
Such an agreement which is unusual for DFT is explained by some
overestimation of structural parameters in the relaxation process.

The valence and conduction gap edge bands
are composed of tellurium and bismuth $p$ states
(Fig.~\ref{fig:structure}(c)), respectively. A strong spin-orbit
interaction leads to a large Bychkov-Rashba type spin splitting of the bulk
states and thus two pairs of extrema along the H-A-L line are
formed: ($E^{\mathrm{H-A}}_{\mathrm{gap}}$) and
($E^{\mathrm{A-L}}_{\mathrm{gap}}$). Under pressure the value of
$E^{\mathrm{H-A}}_{\mathrm{gap}}$ diminishes and at a certain
pressure,$P_c$, shrinks to zero that indicates
normal semiconductor-Weyl semimetal transition.
These doesn't happen with gap along the A-L direction,
$E^{\mathrm{A-L}}_{\mathrm{gap}}$. Upon further increase of
pressure a band gap appears with inverted edges in the vicinity of
the A point: now the lowest conduction band is formed by tellurium $p$
states while the top valence band consists of bismuth $p$ orbitals
(Fig.~\ref{fig:structure}(c)). The calculated value of $P_{c}$ equal
to $\sim$3.73~GPa (Fig.~\ref{fig:structure}(d)) agrees well with the
experimental one, $P_{c}$ = 3.5~GPa.\cite{xi2013,ideue2014}

Due to a small overestimation of the band gap compared to the experimental value ($\sim$7\%), the $P_c$ parameter is also slightly overestimated. The calculated value of $P_c$=3.5~GPa contradicts $P_c$=10~GPa obtained in the $GW$ calculation.\cite{tran2014} As a result according to the Ref.\citenum{tran2014} the $P_c$=10~GPa indicates the impossibility of TPT in BiTeI due to the crystal phase transition at ~9~GPa. The discrepancy is explained by a strong overestimation of the band gap width at zero pressure in the $GW$ calculations. Switching on the corrections on Van der Waals forces lead to a strong underestimation of the band gap at zero pressure (~230 meV). For this reason, we do not present the results obtained after the relaxation of crystal structure with van der Waals corrections.

The behaviour of band gap $E^{\mathrm{H-A}}_{\mathrm{gap}}$ and the
trajectory of crossing (Weyl) points near the topological phase
transition are plotted in Figs.~\ref{fig:structure}(e) and (f).
$E^{\mathrm{min}}_{\mathrm{gap}}$ is the smallest value of
$E^{\mathrm{H-A}}_{\mathrm{gap}}$ in the entire BZ. The position of
Weyl points in the reciprocal space depends on the value of
pressure. The Weyl points are formed at the BZ boundary, $k_{z} =
\pi/c$, along each AH direction. At the phase transition pressure, $P_{c}$,
 each Weyl point splits up into pair of nodes with a clockwise and
counterclockwise propagation on the ($k_{x},k_{y}$) plane
(Fig.~\ref{fig:structure}(f)). Also, these two nodes shift in opposite
directions along $k_{z}$. The Weyl points are observed up to
$\sim$3.93~GPa and then the band degeneracy is lifted and BiTeI
converts into a topological insulator. Thus, the Weyl semimetal
phase exists within the pressure interval from 3.73 to 3.93~GPa. The
range is not small, so the phase can be experimentally observed.

The pairs of Weyl points appeared along the AH directions have
opposite topological charges (chirality) which are defined by Chern
numbers as the flux of Berry phase gauge field over a sphere around
each Weyl point: $C_n=1/(2\pi)\oint \mathbf{\Omega}_n(\mathbf{k})
\mathbf{n} \mathrm{d}S$, where $\mathbf{n}$~-- surface normal vector
and $\mathbf{\Omega}_n(\mathbf{k})=\mathbf{\nabla}_k \times
\mathbf{A_n}(\mathbf{k})$ is the Berry curvature. Here
$\mathbf{A}_n(\mathbf{k})$ is the Berry connection defined as
$\mathbf{A}_n=i \left\langle n\mathbf{k} \right| \mathbf{\nabla}
\left| n\mathbf{k} \right\rangle$ for n-th Bloch state,
 $\left| n\mathbf{k}\right\rangle$,
(in our case~-- the highest occupied band) with quantum number
$\mathbf{k}$. The calculation reveals that the Weyl points which
move clockwise (counterclockwise) have a positive (negative) Chern
number, $C_n$=1 (-1), that corresponds to drain (source) points of
the Berry gauge field. The in-plane components of the Berry
curvature at $k_z=\pi/c-0.012$~\AA$^{-1}$ are demonstrated in
Fig.~\ref{fig:berry}.

\subsection{Surface band structure of BiTeI}

To illustrate surface electronic properties of the systems with Bychkov-Rashba-type spin splitting of bands in the topological and Weyl semimetal phases we considered within tight-binding method a hypothetical model of the bismuth tellurohalides surface under pressure. As was mentioned above to construct the surface under pressure we used large slab with fully relaxed parameters of the bulk structure under the pressure. To do the model more realistic, we included the effect of band bending because of
lack of inversion symmetry in the bismuth tellurohalides.

The topological phase transition is accompanied by qualitative
changes in the surface electronic structure. The BiTeI has two
possible surface terminations: iodine or tellurium with positive and
negative band bending, respectively.~\cite{eremeev_2012,eremeev_jetpl_2012}
Let us first trace the
evolution of surface electronic states with pressure without
considering the effect of band bending (see
Fig.~\ref{fig:surf}(a--f)). In the absence of external pressure
(Figs.~\ref{fig:surf}(a) and (d)), the surface states on the
Te(I)-terminated surface appear along the outer (inner) brunch of
unoccupied spin-orbit split bulk bands and are mainly localized in
the topmost trilayer.

In the pressure-induced Weyl semimetal
phase~(Figs.~\ref{fig:surf}(b) and (e)), the surface states on the
Te-terminated surface remain along the outer edge of the bulk bands
while in the case of iodine termination they shift into the
band gap and exhibit a cone-like dispersion with crossing at the BZ
center. The surface states touch the crossing (Weyl) points of the
conduction and valence bulk bands. In the topological phase
(Figs.~\ref{fig:surf}(c) and (f)), the surface states
become gapless linking the valence and conduction bands. In the case
of I-terminated surface, the crossing of two surface states occurs
in the band gap slightly above the bulk conduction band minima while
on the Te-terminated surface the crossing (Dirac) point is inside off
the valence bands due to the mixing of surface and bulk electronic
bands.

The surface electronic structure obtained with taking into account
the effect of band bending as well as its evolution under pressure
is different. The overall band structure is strongly modified
because a set of well-defined Bychkov-Rashba type spin split surface bands
appearing due
to the band-bending effect.\cite{bahramy2_2012} In the topologically
trivial phase (Figs.~\ref{fig:surf}(g) and (j)), the lowest surface
band on the Te-terminated surface lies within the energy gap and is
strongly localized in the topmost trilayer. The states with higher
energies are confined within
three upper trilayers. On the I-terminated surface, a set of surface
bands splits off from the valence bulk band. The occupied
surface electronic states appear mainly inside or at the edge of the
bulk bands except the topmost state which is shifted upwards into
the energy gap. Such a hierarchy of surface states was also observed
in ARPES experiments.\cite{landolt_2012,sakano_2013} In the Weyl
semimetal phase (Figs.~\ref{fig:surf}(h) and (k)), when passing
through the bulk Weyl nodes the surface states localized in the
upper trilayer become discontinuous. In the topological phase
(Fig.~\ref{fig:surf}(i) and (l)), the surface states are breaking up
into two parts near the bulk edge extrema, and gapless surface
states with a linear dispersion in the band gap are formed. On both
surface terminations, the crossing point of the gapless surface
states is inside the bulk bands where they turn into surface
resonances.

\subsection{Fermi arcs in Weyl-semimetal phase of BiTeI}

The surface electronic structure of Weyl semimetals is characterized by the
presence of Fermi arcs connecting the Weyl points with
opposite chirality. Figure~\ref{fig:fermi_arcs} shows the electronic
spectra at the Fermi level of the (001) surface for both
terminations. The isoenergetic surface spectra were calculated with
and without the account of the band-bending at a pressure of
3.84~GPa which corresponds to a Weyl semimetal phase (see
Fig.~\ref{fig:structure}(e)). Also shown are the positions of Weyl
points on the ($k_{x},k_{y}$) plane. The pairs of Weyl points lie on
the trajectory connecting the positions of Weyl points at the $P_c$ pressure
(3.73~GPa) and at the annihilation point (3.93~GPa) (see
Fig.~\ref{fig:structure}(f)).

Without taking band bending into account the Fermi arcs link up
the pair of Weyl points induced originally along the A-H
($\overline{\Gamma}$--$\overline{\mathrm{K}}$) direction of the bulk
(surface) BZ. The form of Fermi arcs depends on the surface
termination. On the Te-terminated surface, they form arc-like
curves coinciding with the trajectory of a pair of nodes which move
clockwise and counterclockwise over the BZ. In the case of
I-termination, the Fermi arcs connect the pairs of Weyl points by a
straight line. On further increase in pressure, the length of Fermi
arcs increases and when the Weyl semimetal--TI phase transition
occurs the arcs connect with each other forming a gapless
topological surface state.

With the account of band bending, in addition to the Fermi arcs a
circular surface state appears which originates from the outer
branches of Bychkov-Rashba type spin split surface states
(Figs.~\ref{fig:surf}(h) and (k)). In the case of I-termination, the
outer circular state is found farther from the surface BZ center as
compared to the Te-terminated surface. However the isoenergetic
curves for both terminations are qualitatively similar, unlike the
previous case. In addition, with the account of band bending the
Fermi arcs link up the Weyl points from neighbouring pairs. Upon
increasing pressure, the length of Fermi arcs decreases up to zero
at $\sim$3.9~GPa when the transition to topological phase occurs.
After the transition the isoenergetic electronic structure is
presented by a single closed curve of the topological surface state.

The formation of Fermi arcs is closely related to the spin
texture of the surface states. Without account of band bending, the
single closed contour in the topological phase has a clockwise spin
helicity.\cite{bahramy_2012} In the Weyl semimetal phase, Fermi arcs
inherit this behaviour~--- the helicity also has a clockwise
character. The spin texture "connects" the source and drain Weyl
points in the clockwise order, so the Fermi arcs link up the pair of
points emerged along the A-H direction.

Subject to the band bending, the situation is opposite. The Fermi
arcs get the spin texture from the inner closed contour of
Bychkov-Rashba type spin split surface states in the trivial phase which
have a counterclockwise helicity.\cite{eremeev_jetpl_2012} The spin
texture "connects" the source and drain Weyl points in
counterclockwise order, so the Weyl points from neighbouring pairs
are connected.

Thus, in the calculations the modification of the surface state
under pressure depends on whether the band-bending effect is taken
into account or not. The modifications are closely related to the
alteration of the Fermi arcs which characterize the isoenergetic spectra
of Weyl semimetals. Our additional calculations revealed that this picture does not change qualitatively with a strong modification of band bending potential. 
Also, all prognosticated peculiarities of the BiTeI surface electronic structure in topological phase can be revealed in other noncentrosymmetric
alloys where the appearance of the topological or trivial phases can depends, for instance, on the  concentrations. Note, such systems were proposed in Ref.~\citenum{vanderbilt2014} and shown that in cases LaBi$_{1-x}$Sb$_{x}$Te$_{3}$, LuBi$_{1-x}$Sb$_{x}$Te$_{3}$ the topological phase can be induced by variation of the Sb concentration, $x$.

\begin{figure}[t]
\centering
\includegraphics[width=\columnwidth]{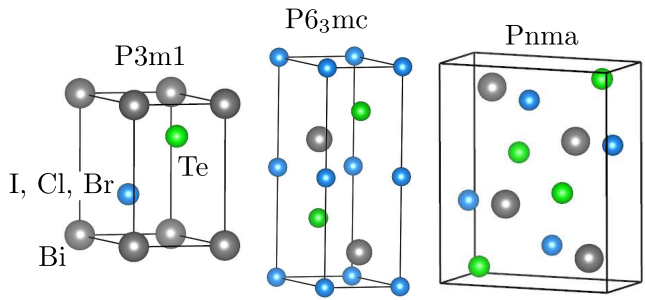}
\caption{Crystal structures of bismuth tellurohalides. All
bismuth-based compounds under consideration were calculated in the
hexagonal P3m1 (left image) and orthorhombic Pnma (right image)
structures. BiTeCl and BiTeI were also considered in the hexagonal
P6$_3$mc phase (central image) which is the structure of BiTeCl at
ambient pressure.} \label{fig:structures}
\end{figure}
\begin{figure}[t!]
\centering
\includegraphics[width=\columnwidth]{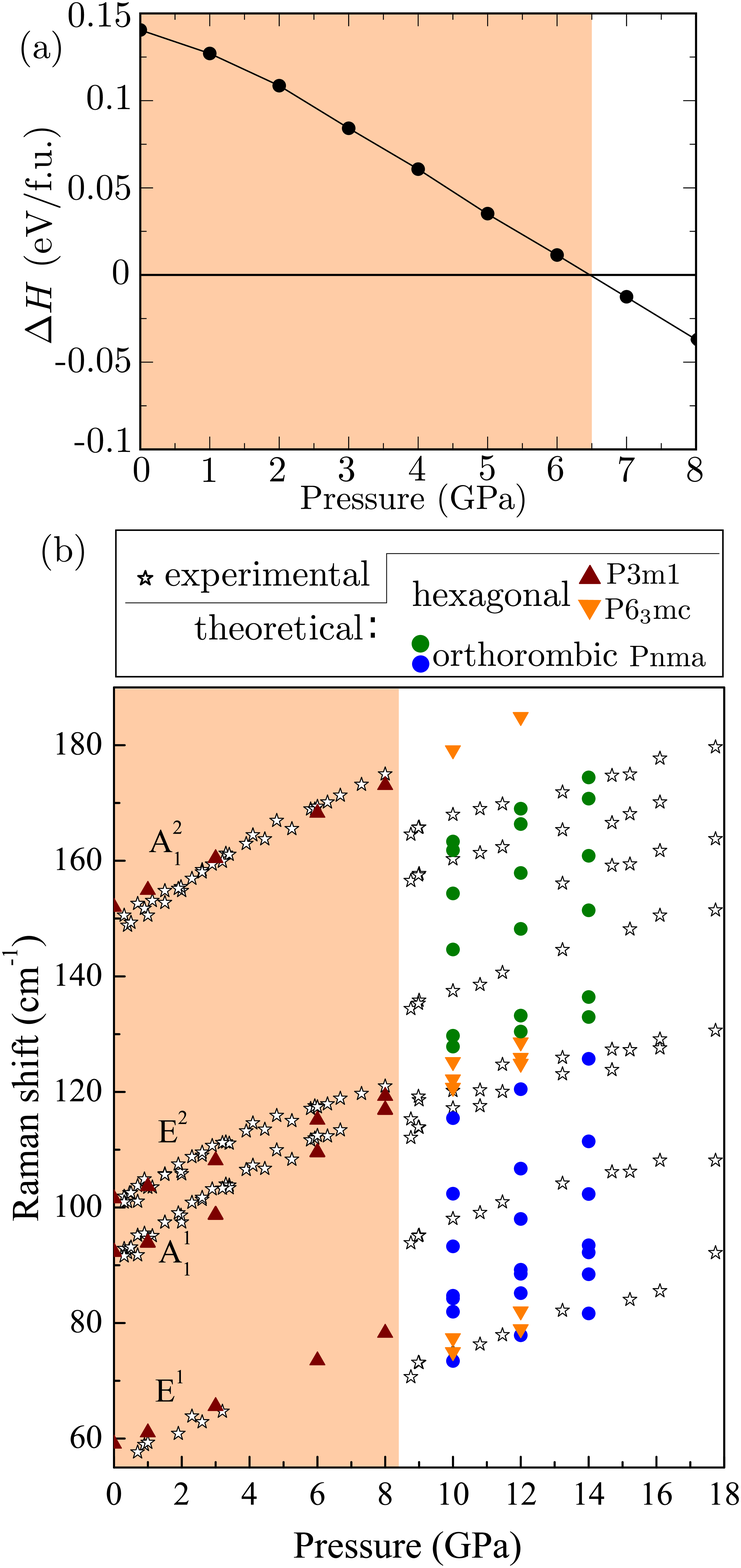}
\caption{(a) Enthalpy per formula unit (f.u.) of the Pnma structure relative to
that of the P3m1 one  ($\Delta H$). (b) Raman mode frequencies
of BiTeI as a function of pressure. Stars denote experimental
data,\cite{ponosov2013,tran2014} triangles (inverted triangles)
and circles show the
values obtained for BiTeI in the hexagonal P3m1 (P6$_3$mc)
and orthorhombic (Pnma) structures, respectively, in the
calculation. In the case of Pnma, blue (green) circles
indicate the modes which involve mostly vibrations of I (Te) atoms.
The pressure range where the P3m1 crystal phase exists is marked by
orange color.} \label{fig:raman}
\end{figure}

\begin{figure*}[t]
\centering
\includegraphics[width=2\columnwidth]{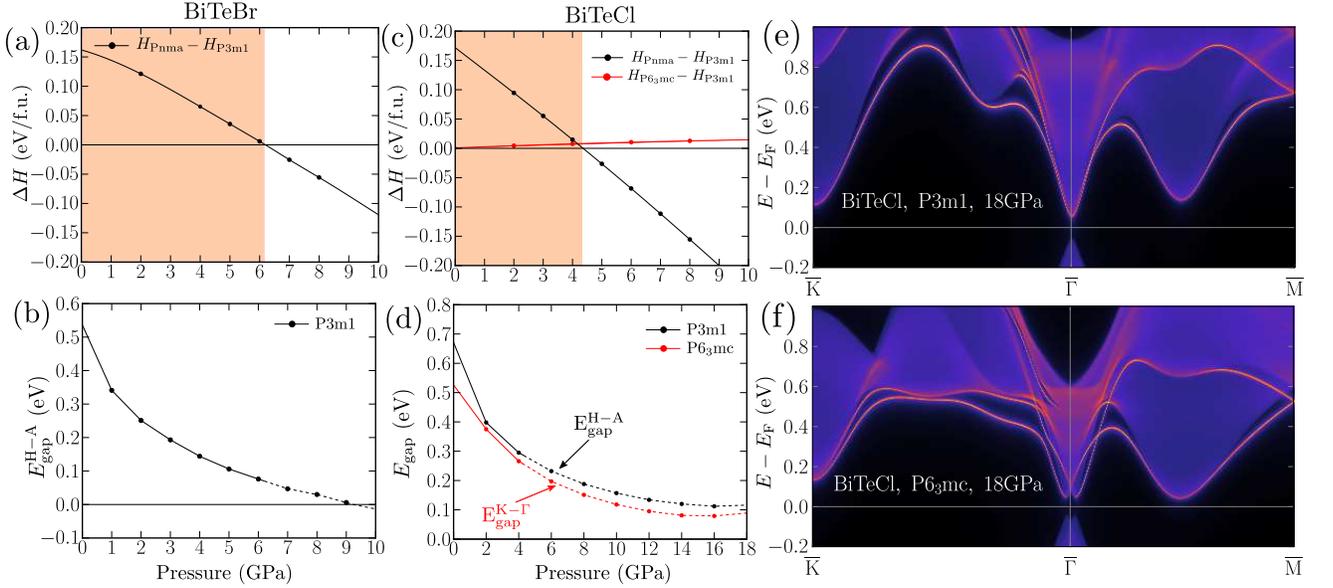}
\caption{
(a) Enthalpy per formula unit (f.u.) of the Pnma structure relative to
that of the P3m1 one  ($\Delta H$). (b) The value of band gap
along the H-A
direction as a function of pressure. (c-d) The same as (a-b) for
BiTeCl but with an additional phase, P6$_3$mc, denoted by red
color. The pressure range corresponding to the low-energy crystal phase
is marked by orange color. The data obtained without taking into account
the crystal phase transition are shown by dashed lines. (e-f)
Surface electronic structure of BiTeCl at a pressure of 18~GPa in
the P3m1 and P6$_3$mc phases, respectively.} \label{fig:btb_btc}
\end{figure*}

\subsection{Crystal phase transition in BiTeI }

Since BiTeI undergoes crystal phase transitions (CPT) under
hydrostatic compression~\cite{xi2013,chen_JPCC,ponosov2013,tran2014}
it is necessary to compare
the pressures corresponding to the TPT and CPT. The X-Ray
diffraction data\cite{xi2013,chen_JPCC} showed that BiTeI remained
in the P3m1 (ambient pressure) structure up to 8--9~GPa. The
stability of the hexagonal phase up to $\sim$9~GPa was also
confirmed by high-pressure Raman spectra
measurements.\cite{ponosov2013,tran2014}
For the higher pressure phase
the orthorhombic Pnma structure with 12 atoms per unit cell  were
suggested (see Fig.~\ref{fig:structures}).\cite{chen_JPCC}
As our calculation show,
this structure is normal non-direct gap semiconductor. We
calculated the difference in enthalpy between the hexagonal P3m1 and
orthorhombic Pnma phases depending on pressure
(Fig.~\ref{fig:raman}(a)). As follows from the calculation the CPT
should occur at $\sim$6.5~GPa. This value is somewhat smaller than the
experimental CPT pressure. Despite this discrepancy, the CPT takes place after the
topological phase transition and at pressures which are beyond the
pressure range (3.73--3.93~GPa) where the Weyl semimetal phase
exists. Thus the CPT is accompanied by a TPT from a topological
insulator to a normal semiconducting phase.

To additionally verify the proposed structure we have also calculated Raman mode
frequencies for the P3m1 phase up to 8~GPa and for the Pnma
structure at pressures 10, 12, and 14~GPa (Fig.~\ref{fig:raman}(b)). In the
hexagonal P3m1 phase with 3 atoms per unit cell, there are two
A$_{1}$ and two E (twofold degenerate) zone center optical modes,
which are both Raman and infrared active because of the lack of
inversion symmetry. At pressures below the CPT, the theoretical data
are in excellent agreement with experimental results with the exception of mode
E$^{1}$ between 4--8~GPa where experimental data are absent
apparently due to decreasing of the mode intensity. In the
experimental Raman spectra\cite{ponosov2013,tran2014} three modes
are visible up to 8~GPa with an expected increase upwards with
pressure. A sudden change in the number and frequency of Raman
active modes which points out to a structural transition is
observed at $\sim$9~GPa.

The Raman spectrum for the orthorhombic phase
has a rather complicated character. All the modes are non
degenerate. Some of them coincide with the experimental data. In general,
variation of Raman mode frequencies with pressure
(10--12--14~GPa) resembles the behaviour of the experimental Raman
modes. However, the number of the calculated Raman modes in the
Pnma structure (13 are shown in the figure while the total number of Raman modes is
equal to 18) is almost twice as large as the experimental data.
Although the experimental Raman spectrum at high pressures bears
some similarity to the calculated Pnma spectrum the measured modes apparently
correspond to a more symmetric structure. The character of the measured
spectrum indicates some orthorhombic phase with similar enthalpy.

Another candidate\cite{tran2014} can be the hexagonal P6$_{3}m$c
structure with 6 atoms per unit cell which is the structure of
another compound, BiTeCl, at ambient pressure.\cite{shevelkov} The
Raman active modes calculated for BiTeI in the hexagonal P6$_{3}m$c
structure at 10 and 12~GPa are continuations of the corresponding
modes in the low-pressure phase. The number of E modes is
doubled and the two lower E modes are shifted down in energy
compared to the E$^{1}$. So the Raman spectrum does not reproduce
the high-pressure experimental data and the P6$_{3}m$c structure
cannot be considered as a high-pressure phase of BiTeI.

\subsection{Electronic properties of BiTeBr and BiTeCl}

For BiTeBr and BiTeCl there is no experimental information about the
possibility of topological phase transition under pressure.
It was thought that
BiTeBr crystallizes in the hexagonal structure, P$\overline{3}$m1,
with Te and Br atoms statistically distributed within two layers
adjacent to the Bi layer.\cite{shevelkov} Later an ordered stacking
of Te and Br sublattices, like in BiTeI, was confirmed by x-ray
diffraction measurements and, moreover, ARPES images revealed a
well-defined Bychkov-Rashba type spin split states.\cite{sakano_2013} The
structure of BiTeCl at ambient pressure is hexagonal, P6$_{3}m$c,
with 6 atoms per unit cell.\cite{shevelkov} We also considered
BiTeCl in the P3m1 crystal phase which was found in the calculation
to be energy preferable up to the CPT. To evaluate the
probability of CPT we also examine both compounds in the
orthorhombic Pnma structure which was originally proposed as a
high-pressure structure for BiTeI.\cite{chen_JPCC} This phase is
considered for BiTeBr and BiTeCl compounds because most V-VI-VII
semiconductors crystallize in this
structure.\cite{donges_1950,donges_1950_2,donges_1951,kikuchi_1967}

As follows from Fig.~\ref{fig:btb_btc}(a) the CPT in BiTeBr takes
place at a pressure of $\sim$6.3~GPa. Although the band gap
(Fig.~\ref{fig:btb_btc}(b)) decreases with pressure its lowest value is of
$\sim$80~meV at the point of CPT and therefore the system
possesses a trivial topology. The CPT pressure is in good
agreement with experimental results of Ref.~\cite{manjon_2016}. The
data exclude the possibility of TPT in the low-pressure phase of
BiTeBr. Note, if another high-pressure structure which is energy
preferable compared to the P3m1 and Pnma structures is realized then
the pressure of CPT would be lower than 6.3~GPa and in any case the
CPT occurs before a TPT.

Due to relatively closer the pressures of TPT and CPT and lacking of
experimental investigations of both transitions, additional calculations
have been provided by inclusion van der Waals correction within DFT-D3~\cite{DFT-D3} scheme.
So, the correction lead to shifting CPT and TPT in the area of smallest pressures:
$\sim$4~GPa and $\sim$8~GPa, respectively. However, the conclusion is preserved: in BiTeBr
TPT is absent.

The case of BiTeCl is more complicated from the crystallographic
point of view. First of all, our results indicate the possibility of
CPT from P6$_3$mc to the P3m1 structure at very small pressures
because the calculated difference in enthalpy
(Fig.~\ref{fig:btb_btc}(c)), red line) turns out to be
$\sim$0.8~meV even at zero pressure. With increasing pressure, the
P3m1 phase remains energy preferable up to 4.3~GPa when the CPT to
the Pnma structure takes place.

However, these results do not affect the general conclusion about
a possibility of TPT in this compound. A TPT does not occur either
in the P3m1 or in the P6$_3$mc crystal phases because although the
band gap width decreases with pressure (Fig.~\ref{fig:btb_btc}(d))
this decreasing practically ceases at a value of $\sim$100~meV
($\sim$80~meV) for the P3m1 (P6$_3$mc) structure. A relatively large
band gap width makes impossible the TPT both in the P3m1
and in the P6$_3$mc hexagonal phase.

This conclusion contradicts the data reported in
Ref.~\onlinecite{chen_Nature} where by APRES measurements a
topological insulator phase in BiTeCl with Dirac-type surface states
was found at zero pressure. The observation of the TI phase might be
explained by a possible crystallization of the material in another
hexagonal structure what is confirmed by a very small bulk band gap width
($\sim$220~meV) obtained in the experiment. The latter contradicts
the electronic structure calculations within the GW-approximation
that reveal the band gap width of 800-900~meV.\cite{rusinov_2013}.

For additional verification of the trivial character of BiTeCl in
the hexagonal phases, we have calculated the surface electronic
structure at a pressure of 18~GPa with taking into account the band
bending effect. Figures~\ref{fig:btb_btc}(e) and (f) show the
electronic spectra of the Te-terminated surface for the P3m1 and
P6$_3$mc crystal phases, respectively. It is obvious that there are
no topological surface states in the bulk band gap and therefore the
system is a trivial semiconductor. Both crystal phases are only
characterized by the presence of well known Bychkov-Rashba type split
surface states. It should be noted, that additional extrema of the
conduction band appear, one at the $\overline{K}$ point and the
other in the $\overline{\Gamma}$-$\overline{M}$ direction which
results in an indirect band gap in the case of the P6$_3$mc
structure.

\section{CONCLUSIONS}

We have investigated a possibility of topological phase transitions
in bismuth tellurohalides. For BiTeI, our results support the
pressure-induced TPT in agreement with experimental
data.\cite{xi2013,chen_JPCC,ideue2014}. The study of bulk and
surface electronic structures of BiTeI under pressure revealed that
an intermediate phase, a Weyl semimetal, is formed during the
transition from a trivial semimetal to a topological insulator in
the pressure interval of $\sim$3.7--3.9~GPa. The range is not too small,
so the intermediate phase can be experimentally observed. The
inclusion of the band bending effect in the calculation allowed us
to consider surface modifications caused by the polarity of BiTeI.
As a result, we revealed the mechanism of changing of the surface
states from a Bychkov-Rashba type spin split state to a gapless surface
state that occurs during the TPT. In the TI phase, the surface
states feature a linear dispersion which depends on the surface
termination. This result is relevant for all asymmetric materials
where TPT can be induced not only by applying hydrostatic pressure
but also by a doping.

We have also compared the Raman mode frequencies obtained
experimentally for BiTeI under pressure with those calculated for
BiTeI. At low pressures the theoretical data for the P3m1 crystal
phase reproduce well the experimental measurements. As for the
high-pressure phases, the experimental Raman spectrum bears some
similarity to the calculated spectrum in the Pnma phase but the measured modes apparently
correspond to a more symmetric structure. We also showed the absence
of TPT in BITeBr and BiTeCl. In BiTeBr, the CPT occurs before a TPT
becomes possible. In BiTeCl, a relatively large value of band gap
width prevents the possibility of TPT up to 18~GPa.

\section{ACKNOWLEDGEMENTS}

This study (research grant No.~8.1.05.2015) was supported by the
Tomsk State University Academic D.I. Mendeleev Fund Program in 2015,
by grant of Saint-Petersburg State University for scientific
investigations No.~15.61.202.2015; the Spanish Ministry of Economy and
Competitiveness MINECO Project FIS2013-48286-C2-1-P.

\vfill\eject


\begin{thebibliography}{9}
\bibitem{pankratov_1987} O.A.~Pankratov, S.V.~Pakhomov and B.A.~Volkov, Solid State Communications {\bf 61}, 93--96 (1987).
\bibitem{fu_2007} L.~Fu, C.L.~Kane and E.J.~Mele, Phys. Rev. Lett. {\bf 98}, 106803 (2007).
\bibitem{xia_2009} Y.~Xia, D.~Qian, D.~Hsieh, L.~Wray, A.~Pal, H.~Lin, A.~Bansil, D.~Grauer, Y.S.~Hor, R.J.~Cava, M.Z.~Hasan, Nature Phys. {\bf 5}, 398 (2009).
\bibitem{eremeev_2010} S.V.~Eremeev, Yu.M.~Koroteev, E.V.~Chulkov, JETP Lett. {\bf 91}, 387-391 (2010).
\bibitem{eremeev_landolt_2012} S.V.~Eremeev, G.~Landolt, T.V.~Menshchikova, B.~Slomski, Yu.M.~Koroteev, Z.S.~Aliev, M.B.~Babanly, J.~Henk, A.~Ernst, L.~Patthey, A.~Eich, A.A.~Khajetoorians, J.~Hagemeister, O.~Pietzsch, J.~Wiebe, R.~Wiesendanger, P.M.~Echenique, S.S.~Tsirkin, I.R.~Amiraslanov, J.H.~Dil and E.V. Chulkov. Nature Communications {\bf 3}, 635 (2012).
\bibitem{BR} Y.A.~Bychkov and E.I.~Rashba, JETP Lett. {\bf 39}, 78 (1984).
\bibitem{ishizaka_2011} K.~Ishizaka, M.S.~Bahramy, M.~Murakawa, M.~Sakano, T.~Shimojima, T.Sonobe, K.~Koizumi, S.~Shin, M.~Miyahara, A.~Kimura, M.Miyamoto, T.~Okuda, H.~Namatame, M.~Taniguchi, R.~Arita, N.~Nagaosa, K.~Kobayashi, Y.~Murakami, R.~Kumai, Y.~Kaneko, Y.~Onose and T.~Tokura, Nature Materials {\bf 10}, 521--526 (2011).
\bibitem{bahramy_2011} M.S.~Bahramy, R.~Arita and N.~Nagaosa, Phys. Rev. B {\bf 84}, 041202 (2011).
\bibitem{eremeev_2012} S.V.~Eremeev, I.A.~Nechaev, Yu.M.~Koroteev, P.M.~Echenique and E.V.~Chulkov, Phys. Rev. Lett. {\bf 108}, 246802 (2012).
\bibitem{eremeev_jetpl_2012} S.V.~Eremeev, I.A.~Nechaev and E.V.~Chulkov, JETP Lett. 96, 437 (2012).
\bibitem{landolt_2012} G.~Landolt, S.V.~Eremeev, Yu.M.~Koroteev, B.~Slomski, S.~Muff, T.~Neupert, M.~Kobayashi, V.N.~Strocov, T.~Schmitt, Z.S.~Aliev, M.B.~Babanly, I.R.~Amiraslanov, E.V.~Chulkov, J.~Osterwalder and J.H.~Dil, Phys. Rev. Lett. {\bf 109}, 116403 (2012).
\bibitem{crepaldi_2012} A.~Crepaldi, L.~Moreschini, G.~Aut\`es, C.~Tournier-Colletta, S.~Moser, N.~Virk, H.~Berger, Ph.~Bugnon, Y.J.~Chang, K.~Kern, A.~Bostwick, E.~Rotenberg, O.V.~Yazyev and M.~Grioni, Phys. Rev. Lett. {\bf 109}, 096803 (2012).
\bibitem{rusinov_2013} I.P.~Rusinov, I.A.~Nechaev, S.V.~Eremeev, C.~Friedrich, S.~Bl\'{u}gel and E. V. Chulkov, Phys. Rev. B. {\bf 87}, 205103 (2013).
\bibitem{sakano_2013} M.~Sakano, M.S.~Bahramy, A.~Katayama, T.~Shimojima, H.~Murakawa, Y.~Kaneko, W.~Malaeb, S.~Shin, K.~Ono, H.~Kumigashira, R.~Arita, N.~Nagaosa, H.Y.~Hwang, Y.~Tokura and K.~Ishizaka, Phys. Rev. Lett. {\bf 110}, 107204 (2013).
\bibitem{eremeev_bitebr} S.V.~Eremeev, I.P.~Rusinov, I.A.~Nechaev and E.V.~Chulkov, New J. Phys. {\bf 15}, 075015 (2013).
\bibitem{dolcini} F.~Dolcini, Phys. Rev. B {\bf 83}, 165304 (2011).
\bibitem{krueckl} V.~Krueckl and K.~Richter, Phys. Rev. Lett {\bf 107}, 086803 (2011).
\bibitem{datta-das} S.~Datta and B.~Das, Appl. Phys. Lett. {\bf 56}, 665 (1990).
\bibitem{datta-das2} J.C. Egues, G.~Burkard and D.~Loss, Appl. Phys. Lett {\bf 82}, 2658 (2003).
\bibitem{bahramy2_2012} M.S.~Bahramy, P.D.C.~King, A. de la Torre, J.~Chang, M.~Shi, L.~Patthey. G.~Balakrishnan, Ph.~Hofmann, R.~Arita, N.~Nagaosa and F.~Baumberger, Nat. Commun. {\bf 3}, 1159 (2012).
\bibitem{bahramy_2012} M.S.~Bahramy, B.J.~Yang, R.~Arita and N.~Nagaosa, Nat. Commun. {\bf 3}, 679 (2012).
\bibitem{xi2013} X.~Xi, C.~Ma, Z.~Liu, Z.~Chen, W.~Ku, H.~Berger, C.~Martin, D.B.~Tanner and G.~L.~Carr, Phys. Rev. Lett. {\bf 111},155701 (2013).
\bibitem{ideue2014} T.~Ideue, J.G.~Checkelsky, M.S.~Bahramy, H.~Murakawa, Y.~Kaneko, N.~Nagaosa and Y.~Tokura, Phys. Rev. B. {\bf 90}, 161107 (2014).
\bibitem{chen_JPCC} Y.~Chen, X.~Xi, W.-L.~Yim, F.~Peng, Y.~Wang, H.~Wang, Y.~Ma, G.~Liu, C.~Sun, C.~Ma, Z.~Chen and H.~Berger, J. Phys. Chem. C. {\bf 117}, 25677--25683 (2013).
\bibitem{vanderbilt2014} J.~Liu and D.~Vanderbilt, Phys. Rev. B. {\bf 90}, 155316 (2014).
\bibitem{ponosov2013} Yu.S.~Ponosov, T.V.~Kuznetsova, O.E.~Tereshchenko, K.A.~Kokh and E.V.~Chulkov, JETP Lett. {\bf 98}, 557--561 (2013).
\bibitem{tran2014} M.K.~Tran, J.~Levallois, P.~Lerch, J.~Teyssier, A.B.~Kuzmenko, G.~Aut\`{e}s, O.V.~Yazyev, A.~Ubaldini, E.~Giannini, D.~van~der~Marel and A.~Akrap, Phys. Rev. Lett. {\bf 112}, 047402 (2014).
\bibitem{manjon_2016} J.A.~Sans,  F.J.~Manj\'on, A.L.J.~Pereira, R.~Vilaplana, O.~Gomis, A.~Segura, A.~Mun\~oz, P.~Rodr\'iguez-Hern\'andez, C.~Popescu, C.~Drasar, and P.~Ruleova, Phys. Rev. B. {\bf 93}, 024110 (2016).
\bibitem{pbe} J.P.~Perdew, K.~Burke and M.~Ernzerhof, Phys. Rev. Lett. {\bf77}, 3865 (1996).
\bibitem{marzari} N.~Marzari and D.~Vanderbilt, Phys. Rev. B {\bf 56}, 12847 (1997).
\bibitem{WZhang} W.~Zhang, R.~Yu, H.-J.~Zhang, X.~Dai and Z.~Fang, New J. Phys. {\bf 12}, 065013 (2010).
\bibitem{Sancho:84} M.P.~Lopez Sancho, J.M.~Lopez Sancho, and J.~Rubio, J. Phys. F {\bf 14}, 1205 (1984).
\bibitem{Sancho:85} M.P.~Lopez Sancho, J.M.~Lopez Sancho, and J.~Rubio, J. Phys. F {\bf 15}, 851 (1985).
\bibitem{Henk:93} J.~Henk and W.~Schattke, Comput. Phys. Commun. {\bf 77}, 69 (1993).
\bibitem{zein} N. E. Zein, Fiz. Tverd. Tela (Leningrad) {\bf 26}, 3028 (1984); N. E. Zein, Sov. Phys. Solid State {\bf 26}, 1825 (1984).
\bibitem{Louie:79} S. G. Louie, K. ­M. Ho, and M. L. Cohen, Phys. Rev. B {\bf 19}, 1774 (1979).
\bibitem{Meyer} B. Meyer, C. Els\"{a}sser, F. Lechermann, and M. F\"{a}hnle, FORTRAN90, Program for Mixed­Basis­Pseudopotential Calculations for Crystals (Max­Planck­Institut f\"{u}r Metallforschung, Stuttgart).
\bibitem{Heid:99} R. Heid and K. ­P. Bohnen, Phys. Rev. B {\bf 60}, R3709 (1999).
\bibitem{Heid:10} R. Heid, K. -P. Bohnen, I. Yu. Sklyadneva, and E. V. Chulkov, Phys. Rev. B {\bf 81}, 174527 (2010).
\bibitem{Sklyadneva} I.Yu. Sklyadneva, R. Heid, K.-P. Bohnen, V. Chis, V. A. Volodin, K. A. Kokh, O. E. Tereshchenko, P. M. Echenique, and E. V. Chulkov, Phys. Rev. B {\bf 86}, 094302 (2012).
\bibitem{shevelkov} A.V.~Shevelkov, E.V.~Dikarev, R.V.~Shpanchenko and B.A.~Popovkin, J. Solid State Chem. {\bf 114}, 379 (1995).
\bibitem{donges_1950} E.~D\"onges, Z. Anorg. Allg. Chem. {\bf 263}, 112 (1950).
\bibitem{donges_1950_2} E.~D\"onges, Z. Anorg. Allg. Chem. {\bf 263}, 280 (1950).
\bibitem{donges_1951} E.~D\"onges, Z. Anorg. Allg. Chem. {\bf 263}, 56 (1951).
\bibitem{kikuchi_1967} A.~Kikuchi, Y.~Oka and E.~Sawaguchi, J. Phys. Soc. Jpn. {\bf 23}, 337 (1967).
\bibitem{DFT-D3} S. Grimme et all., J. Chem. Phys. {\bf 132}, 154104 (2010).
\bibitem{chen_Nature} Y.L.~Chen, M.~Kanou, Z.K.~Liu, H.J.~Zhang, J.A.~Sobota, D.~Leuenberger, S.K.~Mo, B.~Zhou, S-L.~Yang, P.S.~Kirchmann, D.H.~Lu, R.G.~Moore, Z.~Hussain, Z.X.~Shen, X.L.~Qi and T.~Sasagawa, Nature Physics {\bf 9}, 704--708 (2013).





\end{thebibliography}
\end{document}